\documentclass[10pt,twocolumn]{revtex4}
\usepackage[latin9]{inputenc}
\usepackage[letterpaper]{geometry}
\usepackage{amstext}
\usepackage{amssymb}
\usepackage{graphicx}

\makeatletter
\@ifundefined{textcolor}{}
{%
 \definecolor{BLACK}{gray}{0}
 \definecolor{WHITE}{gray}{1}
 \definecolor{RED}{rgb}{1,0,0}
 \definecolor{GREEN}{rgb}{0,1,0}
 \definecolor{BLUE}{rgb}{0,0,1}
 \definecolor{CYAN}{cmyk}{1,0,0,0}
 \definecolor{MAGENTA}{cmyk}{0,1,0,0}
 \definecolor{YELLOW}{cmyk}{0,0,1,0}
}

\makeatother

\begin{document}

\title{Excluded volume effects in polymer brushes at moderate chain stretching}

\author{Dirk Romeis and Michael Lang}

\address{Leibniz-Institut f\"ur Polymerforschung Dresden e.V., Hohe Strasse
6, 01069 Dresden, Germany.}
\begin{abstract}
We develop a strong stretching approximation for a polymer brush made
of self-avoiding polymer chains. The density profile of the brush
and the distribution of the end monomer positions in stretching direction
are computed and compared with simulation data. We find that our approach
leads to a clearly better approximation as compared to previous approaches
based upon Gaussian elasticity at low grafting densities (moderate
chain stretching), for which corrections due to finite extensibility
can be ignored.
\end{abstract}
\maketitle

\section{Introduction}

Only ideal chain models have been assumed in previous literature for
the analytic self-consistent calculation of the polymer brush density
profile and the distribution of free ends. Within such an approach
\cite{MWC,ZhuPriBo}, the Gaussian elasticity is typically balanced
by repulsive two body interactions, similar to the Flory theory \cite{Flory}
of a polymer in good solvent. This analytic framework is known as
the 'Strong Stretching Approximation' (SSA), since it also neglects
fluctuation effects. It was later extended \cite{ShimCates,LaiHalperin,AmoskovPryam,BieshAmoskov}
to account for finite chain extensibility whereby it was shown recently
that finite extensibility enhances the surface instability effect
in dense brushes \cite{Dirk 1,Dirk 2}.

The SSA predictions for the monomer density profile $\phi(z)$ and
the distribution of free ends $P_{N}(z)$ in good solvent have been
tested widely with simulation data of brushes made of self-avoiding
polymer chains \cite{MuratGrest,ChakTor,DickHong,LaiBinder1,LaiBinder2,ChenFwu,HeMerlitz,KaraBit}.
The 'standard' SSA results \cite{MWC,ZhuPriBo} were found to give
a far better agreement to the simulation data as compared to the step-like
profile approach suggested by Alexander and de Gennes \cite{AlexanderBrush,deGennesBrush}.
Nevertheless, qualitative differences to the SSA approximation were
detected \cite{MuratGrest,ChakTor,DickHong,LaiBinder1,LaiBinder2,ChenFwu,HeMerlitz,KaraBit}.
In particular, deviations were found for the distribution of free
ends, which were often asigned to finite extensibility effects \cite{DickHong,ChenFwu,HeMerlitz,BieshAmoskov},
even though a rather low grafting density $\sigma$ was reached in
some of these works. For a better understanding of this point, we
focus in the present work on the limit of low grafting densities and
moderate chain stretching.

As pointed out by de Gennes \cite{deGennes_Book}, the Flory theory
leads to large errors for the individual contributions of elasticity
and two body interactions, while the average chain extension is approximated
well in most cases. This fact is reflected by the estimate for the
chain conformations in a polymer brush, since SSA using ideal chains
\cite{MWC,ZhuPriBo} and scaling models \cite{AlexanderBrush,deGennesBrush}
predict essentially the same scaling of brush height as function of
the uniform grafting density $\sigma$ and the monodisperse degree
of polymerization of the chains $N$. However, significant deviations
can be expected, if only one of the contributions to the Flory theory
plays a major role for a particular feature of the brush.

For the above self-consistent models \cite{MWC,ZhuPriBo,ShimCates,LaiHalperin,AmoskovPryam,BieshAmoskov},
the static fluctuations of individual monomers around their average
positions are dominated by the elasticity of the chains. Both, scaling
models \cite{AlexanderBrush,deGennesBrush} and SSA \cite{MWC,ZhuPriBo}
assume that the chain conformations can be approximated as a string
of blobs of suitable sizes that is fully stretched in $z$-direction
perpendicular to the grafting plane. At equilibrium conformations,
there is a local balance of the Pincus blob \cite{Pincus} (due to
tension) and the excluded volume interactions with neighboring ``blobs''\cite{deGennes_Book}.
Any fluctuation in $z$-direction that causes an extension of the
chain size beyond its average length leads to Pincus blobs smaller
than the excluded volume blobs. Thus, we expect that fluctuations
towards larger chain extensions are probing the deformation behaviour
of self-avoiding polymer chains in contrast to the previously assumed
Gaussian elasticity.

Fluctuations that lead to a reduction of the chain extension, on the
other hand, might be approximated by an almost fully stretched chain
of blobs in a melt of blobs. For a polymer melt (and thus, for the
blobs of a semi-dilute solution of polymers) it was well accepted
that the excluded volume is rather small, which leads to nearly ideal
chain conformations \cite{Flory_Book}. In recent years, it was shown
\cite{Wittmer_2004,Wittmer_2007} that Flory's result is correct in
the thermodynamic limit, but long range correlations along the chains
lead to a slow cross-over from self-avoiding like chain conformations
on the level of consecutive segments (or blobs) to nearly ideal chain
conformations for segments far apart from each other along a chain.
Thus, Ref's \cite{Wittmer_2004,Wittmer_2007} predict implicitly an
elasticity similar to self-avoiding chains, if the tension blobs remain
of similar size than the excluded volume blobs and elasticity is probed
on the level of consecutive blobs. For large fluctuations towards
small chain extensions, the tension blobs increase in size and a continuous
but slow cross-over to Gaussian elasticity is expected for the static
monomer fluctuations, as for instance, the distribution of the free
ends $P_{N}(z)$. Note that this slow cross-over for the static monomer
fluctuations \cite{Lang_2013} was observed previously for unentangled
polymer networks.

In the present work, we want to demonstrate that previously observed
deviations to the classical SSA approximation extend to very low grafting
densities, for which finite extensibility cannot be used as an explanation.
We show that the differences between data and theory arise from the
approximations used in previous self-consistent calculations. We elaborate
this point in the framework of a strong stretching approximation \cite{AmoskovPryam}
that is extended to incorporate both, the elasticity of self-avoiding
chains and the scaling expression for the osmotic pressure (Section
\ref{Model}). The theory is written in general form, such that good
solvent or $\Theta$-solutions are derived simultaneously. The obtained
results are presented in comparison to previous works in section \ref{Results}.
Furthermore, in section \ref{Test}, we put the new predictions to
a test with simulation data on polymer brushes at very low grafting
densities such that effects of finite extensibility can be ignored.
In section \ref{Scaling}, we discuss the scaling of the distribution
of free ends and show that simulation data is in agreement with our
model but not with the classical SSA. Finally, we discuss our results
in section \ref{Discussion}.

\section{The Strong Stretching Approximation for a self-avoiding Chain}

\label{Model}

To simplify notation, we normalize all length units by the length
of a Kuhn segment of the chains such that all lengths become dimensionless.
Similarily, we make the energy scale dimensionless by putting $k_{B}T=1$,
with the Boltzman constant $k_{B}$ and the temperature $T$. Furthermore,
we drop all coefficients of order unity, if these do not arise from
our computations. Below we consider only the model case of a monodisperse
layer of chains with degree of polymerization $N$ that are grafted
at a constant grafting density $\sigma$.

Let us first discuss the limits and simplifications of the strong
stretching approximation. Due to the assumption that the degree of
polymerization tends to infinity, $N\to\infty$, corrections and fluctuations
(which are of order $\approx N^{1/2}$) are fully neglected \cite{Milner_MM89_2}.
Hence, the fluctuating tail at the brush surface as observed in simulations
and numerical calculations is missing \cite{Milner_JCSFT_90}. Also,
the depletion effect within the first blob next to the grafting surface
is ignored. With increasing $N$, the agreement to SSA is enhanced,
as these effects become unimportant.

To apply a SSA, the grafting density $\sigma$ must be large enough
to ensure sufficient overlapping of neighbouring chains. Hence, $\sigma\gg\sigma^{*}$,
with $\sigma^{*}=1/(\pi R_{g0}^{2})$ the overlap density and $R_{g\text{0}}$
denoting the radius of gyration of an isolated chain at same solvent
conditions. In Ref. \cite{KaraBit}, this lower bound is estimated
as $\sigma\geq8/(\pi R_{g0}^{2})$. Since the 'standard' SSA \cite{MWC,ZhuPriBo}
neglects finite extensibility of the chains, $\sigma$ must be bound
from above to ensure that the chains are not 'overstretched'. This
upper bound was roughly estimated as $\sigma\lesssim0.2$ \cite{LaiHalperin,AmoskovPryam}.
In the present work, we also want to neglect finite extensibility
effects and conclude for the limits of our approximation: 
\begin{equation}
\frac{8}{\pi R_{g0}^{2}}\leq\sigma\lesssim0.2\label{validregime}
\end{equation}

The starting point for our approach is a model equation for the chain
deformation. In the limit of weak or moderate chain stretching, all
ideal chain models show a linear elongation-force relation 
\begin{equation}
\frac{r}{N}\propto e(f)\propto f,\label{idealfx}
\end{equation}
where $r/N$ denotes the relative elongation with respect to the contour
length (chain length $N$) and the applied force $f$. Let us write
the elongation-force relation for a self-avoiding chain in general
form \cite{Pincus} 
\begin{equation}
e(f)\propto f^{\frac{1-\nu}{\nu}},\label{pincusfx}
\end{equation}
such that it describes simultaneously a self-avoiding polymer with
Flory exponent $\nu\approx0.588\approx3/5$ \cite{LeGuillou} in good
solvents and an ideal chain or a self-avoiding polymer in $\Theta$-solvents
with $\nu=1/2$. As discussed in the introduction, we assume that
the elasticity of a chain in a brush at good solvent conditions is
well described by an exponent of $\nu\approx0.588$ in constrast to
previous works. Note that the same elongation-force relation was used
to describe polyelectrolyte brushes \cite{Zhulina_PE}, for which
the Pincus blob is smaller than the concentration blob due to the
osmotic pressure of the counterions. In the following, we only consider
the direction perpendicular to the grafting surface (in $z$-direction)
and assume that the brush is homogeneous in lateral $xy$-directions.

In the SSA, it is considered that the elongation-force relation of
Equation (\ref{idealfx}), resp.\ Equation (\ref{pincusfx}), is
'locally' valid \cite{AmoskovPryam}. Let us introduce $s:=i/N$ to
denote the normalized chain contour, where $i$ represents the $i$-th
segment and discuss the elongation-force relation $e(f)$ under a
locally applied force. In this view, the inverse of $e(f)$ gives
the amount of 'segments' $ds$ being localized in the element $dz$
at position $z$. 

Let us choose the boundary conditions such that the first segment
is pinned at $z(s=0)=0$ and the last segment is elongated to position
$z(s=1)=z_{e}$. Then, we can write for the countour integral 
\begin{equation}
\int_{0}^{1}ds=\int_{0}^{z_{e}}\frac{ds}{dz}dz=\int_{0}^{z_{e}}\frac{dz}{e(f)N}=1.\label{contourcond}
\end{equation}
In the following, we describe the chains as a non-harmonic spring
in a non-uniform stretching field. This stretching field along the
'contour' $s$ of the chain is generated by a potential field $\Phi(z)$
in space \cite{AmoskovPryam}: 
\begin{equation}
\frac{\partial f(z(s))}{\partial s}=-\frac{\partial\Phi(z)}{\partial z}.\label{potfield}
\end{equation}
The following boundary conditions must be fulfilled: $z(s=0)=0$ (grafting
condition), $f(s=1)=0$ (no force on the free chain end in equilibrium)
and $z(s=1)=z_{e}$ (position of the free chain end). Integrating
both sides in Equation (\ref{potfield}) under these boundary conditions
we have: 
\begin{equation}
\int_{f(z_{e})=0}^{f(z)}e(f')df'=\Phi(z_{e})-\Phi(z).\label{potential}
\end{equation}
With Equation (\ref{pincusfx}) we find the following relation between
force and potential field 
\begin{equation}
f\propto\left(\frac{\Phi(z_{e})-\Phi(z)}{\nu}\right)^{\nu}.\label{forcepotrel}
\end{equation}

The self-consistent potential must be that one where the spring is
in a state of indifferent equilibrium \cite{AmoskovPryam}. This is
the only possibility to provide a continuous distribution $P_{N}(z_{e})$
of the free ends of monodisperse chains. The self-consistency condition
(or the indifferent equilibrium condition) in case of arbitrary $\nu$
is derived from combining Equation (\ref{contourcond}) with (\ref{pincusfx})
and (\ref{forcepotrel}) 
\begin{equation}
\int_{0}^{z_{e}}\frac{\frac{1}{N}dz'}{\left(\Phi(z_{e})-\Phi(z')\right)^{1-\nu}}=const.\ \forall z_{e}.\label{indiffequ2}
\end{equation}
Using $q=z'/z_{e}$ this transforms to 
\begin{equation}
\int_{0}^{1}\frac{\frac{z_{e}}{N}dq}{\left(\Phi(z_{e})-\Phi(z_{e}q)\right)^{1-\nu}}=const.\ \forall z_{e},\label{indiffequ3}
\end{equation}
which can only become independent of $z_{e}$, if 
\begin{equation}
\Phi(z)\propto\left(\frac{z}{N}\right)^{1/(1-\nu)}.\label{Vze}
\end{equation}
In case of a Gaussian chain (or a chain in $\Theta$-solvent with
$\nu=1/2$) we recover with $e(f)=f/3$, the results of previous works
\cite{MWC,ZhuPriBo,AmoskovPryam,BieshAmoskov} 
\begin{equation}
\Phi(z)=\frac{3\pi^{2}}{8}\left(\frac{z}{N}\right)^{2}\approx3.71\left(\frac{z}{N}\right)^{2},\label{gaussscpot}
\end{equation}
while for a swollen chain we find for $\nu=3/5$ and introducing a
spring constant $a$ in Equation (\ref{pincusfx}) that 
\begin{eqnarray}
\Phi(z) & = & \frac{3}{5}\left(2-\frac{2}{\sqrt{5}}\right)^{5/4}\left(\frac{2\pi}{5}\right)^{5/2}a^{-3/2}\left(\frac{z}{N}\right)^{5/2}\nonumber \\
 & \approx & \frac{1.2}{a^{3/2}}\left(\frac{z}{N}\right)^{5/2}.\label{VSC}
\end{eqnarray}

To calculate the profile of the monomer volume fraction $\phi(z)$
and the free end distribution $P_{N}(z)$ we need to define an equation
of state relating the monomer chemical potential $\mu$ to $\phi(z)$.
For the description of polymer brushes originally \cite{MWC,ZhuPriBo}
only the leading order of the virial expansion of the osmotic pressure
\begin{equation}
\Pi\approx\mbox{v}\phi^{2}\label{meanosmo}
\end{equation}
in good solvent was considered. This is similar to the Flory theory,
where Gaussian elasticity is combined with repulsive two body interactinos.
Here, the Parameter $\mbox{v}$ denotes the excluded volume.

In analogy to scaling estimates for the equilibrium degree of swelling
of a polymer network \cite{Rubinstein}, we combine the scaling approach
for the osmotic pressure of a semi-dilute solution of polymers \cite{Rubinstein}
\begin{equation}
\Pi\approx\mbox{v}^{(6\nu-3)/(3\nu-1)}\phi^{3\nu/(3\nu-1)},\label{osmotic}
\end{equation}
with the scaling of chain elasticity as described by equation (\ref{pincusfx}).
Note that the osmotic pressure is $\propto\phi^{3}$ in $\Theta$-solutions
with $\nu=1/2$ and $\propto\phi^{2.3}$ with $\nu\approx0.588$ for
good solvents.

From Equation (\ref{osmotic}), the monomer chemical potential $\mu$
can be obtained from considering the change in free energy 
\begin{equation}
\Delta F=-\int pdV\propto(3\nu-1)\mbox{v}^{(6\nu-3)/(3\nu-1)}M\phi^{1/(3\nu-1)}\label{freeenergy}
\end{equation}
using $\phi=M/V$. Here, $M$ is the total number of monomers in the
system, $p$ the pressure and $V$ the volume of the system as measured
in unit volumes. The chemical potential is the change in free energy
$F$ upon adding or removing a monomer at constant volume and temperature.
Thus,

\begin{equation}
\mu(\phi)=\left(\frac{\partial F}{\partial M}\right)_{V,T}\propto3\nu\mbox{v}^{(6\nu-3)/(3\nu-1)}\phi^{1/(3\nu-1)}.\label{virial}
\end{equation}
Let us use $H$ to denote the brush height. Since the monomer chemical
potential $\mu(\phi)$ must provide the self-consistent potential
profile we have 
\begin{equation}
\mu(z)=\Delta\Phi(z)=\Phi(H)-\Phi(z)\label{eq:mu(z)}
\end{equation}
and hence 
\begin{equation}
\phi(z)\propto\mbox{v}^{-(6\nu-3)}\left(\frac{\Phi(H)-\Phi(z)}{3\nu}\right)^{3\nu-1},\label{mupot}
\end{equation}
where $\Phi(H)$, and subsequently the brush height $H$, is fixed
by the normalization condition $\int_{0}^{H}\phi(z)dz=N\sigma$. With
Equations (\ref{mupot}) and (\ref{Vze}) we obtain 
\begin{equation}
\int_{0}^{H}\, dz\left(\left(\frac{H}{N}\right)^{\frac{1}{1-\nu}}-\left(\frac{z}{N}\right)^{\frac{1}{1-\nu}}\right)^{3\nu-1}\propto N\sigma.\label{normingcond}
\end{equation}
Using $y:=z/H$ this condition becomes 
\begin{equation}
\int_{0}^{1}\, dy\left(1-y^{1/(1-\nu)}\right)^{3\nu-1}\left(\frac{H}{N}\right)^{(2\nu)/(1-\nu)}\propto\sigma.\label{normingcond2}
\end{equation}
As a consequence of this analysis, we recover the scaling relation
for the brush height \cite{AlexanderBrush,deGennesBrush} 
\begin{equation}
H\propto N\mbox{v}^{(2\nu-1)/\nu}\sigma^{(1-\nu)/2\nu}.\label{height}
\end{equation}
Note, that for $\Theta$-solvents with $\nu=1/2$ the relations assumed
in Equation (\ref{pincusfx}) and (\ref{osmotic}) lead to identical
results as previous work on a polymer brush in $\Theta$-solution
\cite{ZhuPriBo}. Finally, we have to mention that the 'standard'
SSA leads to a prediction $H\propto(\mbox{v}\sigma)^{1/3}N$ \cite{MWC,ZhuPriBo}
that agrees with the scaling result by coincidence, if $\nu=3/5$,
as discussed previously \cite{KreerMetzger}.

\section{The brush profile and the distribution of end monomers}

\label{Results}

The brush profile $\phi(z)$ and the free end distribution $P_{N}(z)$
of the brush chains differ considerably from previous self-consistent
models in the 'standard' SSA approximation of Ref's \cite{MWC,ZhuPriBo}.
The present approach, combining the scaling relations for elongation
and osmotic pressure, predicts a clearly different density profile
(using Equation (\ref{Vze}) in Equation (\ref{mupot})): 
\begin{eqnarray}
\phi(z) & = & \phi_{0}\left(1-\left(\frac{z}{H}\right)^{1/(1-\nu)}\right)^{3\nu-1}\label{densitygeneral}\\
 & \approx & \phi_{0}\left(1-\left(\frac{z}{H}\right)^{2.43}\right)^{0.76}\label{densitypincus}
\end{eqnarray}
in contrast to the 'classical' parabolic profile 
\begin{equation}
\phi(z)=\phi_{0}\left(1-\left(\frac{z}{H}\right)^{2}\right)\label{eq:parabolic}
\end{equation}
of Ref. \cite{MWC,ZhuPriBo}. In the case of $\Theta$-solvent, setting
$\nu=1/2$ in eq.\ (\ref{densitygeneral}), we obtain the same result 

\begin{equation}
\phi(z)=\phi_{0}\left(1-\left(\frac{z}{H}\right)^{2}\right)^{1/2}.\label{eq:Thetaprofil}
\end{equation}
as Ref. \cite{ZhuBoPry} using the 'standard' SSA for ideal chains
in $\Theta$-solvent. 

\begin{figure}
\begin{centering}
\includegraphics[width=1\columnwidth]{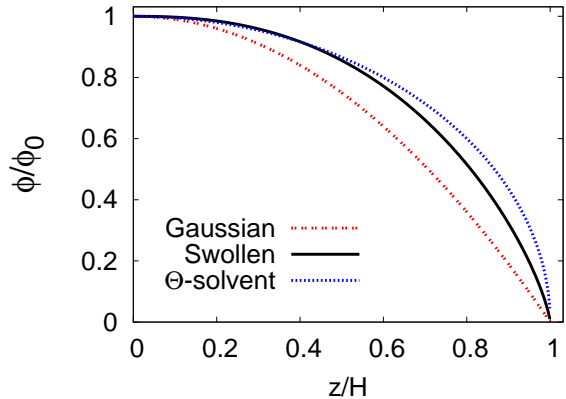}
\par\end{centering}

\caption{\label{figdensity}Comparison of the normalized monomer density profiles
for different approximations: Gaussian elasticity combined with second
virial (Gaussian, eq.\ (\ref{eq:parabolic})), elasticity of self-avoiding
polymers combined with the scaling description of osmotic pressure
(Swollen, eq.\ (\ref{densitypincus})), and Gaussian elasticity combined
with zero second virial and non-zero third virial coefficient ($\Theta$-solvent,
eq.\ (\ref{eq:Thetaprofil})).}
\end{figure}

Note, that the particular brush height $H$ and the maximum monomer
density $\phi_{0}$ depend on $\sigma$ and details of the model as,
for instance, the spring constant in Equation (\ref{pincusfx}) or
the excluded volume parameter $\mbox{v}$ in Equation (\ref{osmotic}).
But the density profile in reduced units $\phi/\phi_{0}$ and $z/H$
leads to a universal plot for each of the above approximations. In
Figure \ref{figdensity}, we compare the normalized monomer density
profiles of our approach with previous models \cite{MWC,ZhuPriBo}.
So far, the 'Strong Stretching Approach' in the low to moderate stretching
approximation (low to moderate $\sigma$) \cite{MWC,ZhuPriBo} always
assumed a Gaussian chain elongation also in good solvent leading to
a harmonic density profile ($\phi\sim-z^{2}$). The $\phi(z)$ in
Equation (\ref{densitypincus}) leads to a flattened density profile
at the grafting surface with a steeper drop at the top of the brush
in comparison to 'standard' SSA calculations \cite{MWC,ZhuPriBo}
making the density profile qualitatively more similar to the $\Theta$-solvent
prediction.

In comparison to the changes for the density profile, the differences
for the free end distribution $P_{N}(z)$ are more pronounced. $P_{N}(z)$
is found \cite{AmoskovPryam} via an implicit integral equation that
becomes 
\begin{equation}
\phi(z)\propto\sigma\int_{z}^{H}\frac{P_{N}(z')dz'}{\left(z'^{1/(1-\nu)}-z^{1/(1-\nu)}\right)^{(1-\nu)}}\label{ge}
\end{equation}
for swollen chains. With Equation (\ref{densitypincus}) and using
$y'=z'/H$, resp.\ $y=z/H$, this leads to 
\begin{equation}
\left(1-y^{\frac{1}{1-\nu}}\right)^{3\nu-1}\propto\int_{y}^{1}\frac{P_{N}(y')dy'}{\left(y'^{\frac{1}{1-\nu}}-y^{\frac{1}{1-\nu}}\right)^{(1-\nu)}}.\label{ge3}
\end{equation}
One can proof by insertion that $P_{N}$ must be of the following
form: 
\begin{eqnarray}
P_{N}(y) & \propto & y^{\nu/(1-\nu)}\left(1-y^{1/(1-\nu)}\right)^{2\nu-1}\label{gegeneral}\\
 & \stackrel{\propto}{\sim} & y^{1.43}\left(1-y^{2.43}\right)^{0.176}.\label{ge2}
\end{eqnarray}

\begin{figure}
\begin{centering}
\includegraphics[width=1\columnwidth]{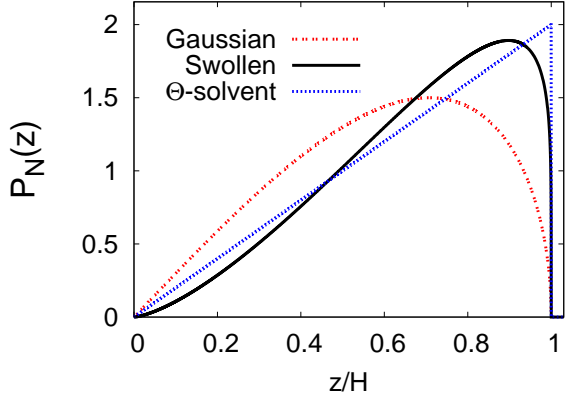}
\par\end{centering}

\caption{\label{figenddistr}Comparing the free end distribution $P_{N}(z)$
for the same models as in Figure \ref{figdensity}.}
\end{figure}

The prediction of Equation (\ref{ge2}) is compared in Figure \ref{figenddistr}
with the predictions from Refs.\ \cite{MWC,ZhuPriBo} for the Gaussian
self-consistent field approach in good solvent: 
\begin{equation}
P_{N}(y)\propto y\left(1-y^{2}\right)^{1/2}\label{gegauss}
\end{equation}
and the solution in $\Theta$-solvent \cite{ZhuBoPry} (also obtained
within the present approach for $\nu=1/2$ in eq.\ (\ref{gegeneral})):
\begin{equation}
P_{N}(y)\propto y.\label{getheta}
\end{equation}
Note, Equations (\ref{ge2})-(\ref{getheta}) only hold for $0\leq y=z/H\leq1$
and are zero otherwise.

The signature of our approximation is the non-linear $z^{\nu/(1-\nu)}$
dependence at small and intermediate $z$ and the clear shift of the
peak position of $P_{N}(z)$ towards the upper free end of the brush
in the case of good solvent. The predictions of the above models are
now compared with simulation data.

\section{Comparison with Simulation Data}

\label{Test}

\begin{figure}
\begin{centering}
\includegraphics[width=1\columnwidth]{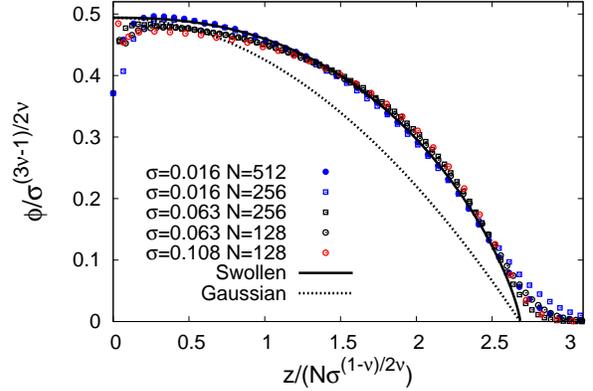}
\par\end{centering}

\caption{\label{figdensitiesBFM}Comparing of Equations (\ref{densitypincus})
(full lines) and (\ref{eq:parabolic}) (dotted lines) with simulation
data of the Bond Fluctuation Model. Monomer density $\phi$ and height
position $z$ are presented in rescaled units according to \cite{AlexanderBrush,deGennesBrush},
$z/H\propto z/(N\sigma^{(1-\nu)/2\nu})$.}
\end{figure}

In a recent work \cite{Lang_Brush}, the cross-linking process of
polymer brushes was studied using the Bond Fluctuation Model (BFM)
of Ref. \cite{Carmesin,DeutschBinder}. The equilibrated and not yet
linked brushes of this work were used as starting basis for sampling
$\phi(z)$ and $P_{N}(z)$ and we refer the reader to this work for
further information on the parameters of the samples, the simulation
method and the equilibration of the brushes. Additional samples with
large $N=256$ and $N=512$ at very low grafting density $\sigma=1/64\approx0.016$
were created in similar manner in order to explore the limit at which
finite extensibility can be ignored; further details of these additional
simulations courtesy in \cite{Lang_Brush2}. To collect data over
an extremely long time interval, we use a GPU implementation of the
BFM \cite{Nedelcu} and equilibrated the brushes for more than one
decade of relaxation times of the brush polymers.

In Figure \ref{figdensitiesBFM}, we display the resulting density
profiles from these simulations in rescaled units. The predicted profile
obtained in the present approach for swollen chains fits quite perfectly
to the simulation data and provides a clear improvement in comparison
with the Gaussian chain approach. In the Figure, we fitted the data
only to our model and plotted the Gaussian chain approach for same
$H$ for comparison. The discrepancies between theory and data can
be devoted to depletion effects at the (hard) grafting surface and
fluctuation effects at the brush surface. For data at largest grafting
densities shown in Figure \ref{figdensitiesBFM}, we observe a weak
tendency towards a more step-like profile that we devote to the onset
of finite extensibility of the chains and the increasing density in
the brush, since the limits of our approximation are approached, see
eq.\ (\ref{validregime}).

\begin{figure}
\begin{centering}
\includegraphics[width=1\columnwidth]{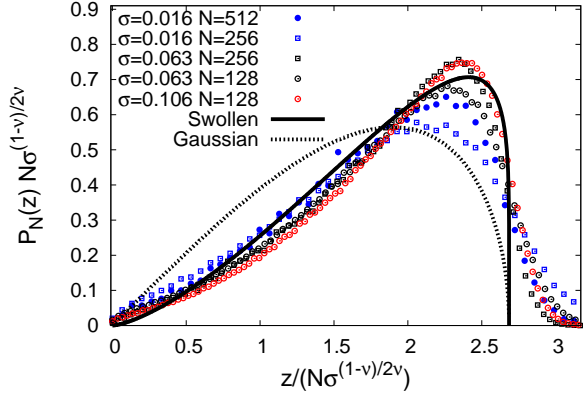}
\par\end{centering}

\caption{\label{figendsBFM}Comparison of the free end distributions $P_{N}(z)$
from BFM simulations at different grafting densities and chain lengths
in rescaled units $z/H\propto z/(N\sigma^{(1-\nu)/2\nu})$, using
the same $H$ as in Figure \ref{figdensitiesBFM}. Normalization:
$\int_{0}^{1}\, d(z/h)\, P_{N}(z/h)=1$.}
\end{figure}

In Figure \ref{figendsBFM}, we test our prediction for the distribution
of free chain ends, eq.\ (\ref{ge2}), with the corresponding simulation
data. Also for $P_{N}(z)$ our approach fits notably better in comparison
to the 'standard' SSA result \cite{MWC,ZhuPriBo}, eq.\ (\ref{eq:parabolic}).
Especially the non-linear slope of $P_{N}(z)$ at small $z$ positions
can be explained using our approximation. 

The trend of a more flattened density profile $\phi(z)$ and a rather
convex (instead of a concave) slope for the distribution $P_{N}(z)$
of free ends is consistent with numerous data presented in previous
works \cite{MuratGrest,ChakTor,DickHong,LaiBinder1,LaiBinder2,ChenFwu,HeMerlitz,KaraBit},
where the simulation results were tested against the 'standard' SSA
predictions. Often such deviations were devoted to finite extensibility
effects \cite{ChenFwu,HeMerlitz}, which is correct for grafting densities
larger as in the present study. But for the low grafting densities
of our work, finite extensibility cannot be used to explain the dicrepancies
to the 'standard' SSA. The above results demonstrate that our model
is a more accurate approximation for polymer brushes of self-avoiding
chains at low grafting densities than previous approaches.

Yet another indication is provided in Refs.\ \cite{LaiBinder1,LaiBinder2},
where the authors compare simulation results in good and in $\Theta$-solvent.
Comparing the chain end distributions for both cases, one can note
a clear change in the slope of $P_{N}(z)$ at small $z$, turning
from convex in good solvent to a rather linear slope in $\Theta$-solvent,
as predicted within our approach (see Figure \ref{figdensity}). But
the most convincing proof in favor of our approximation is presented
in the following section.

\section{A test of the scaling of the distribution of free ends}

\label{Scaling}

\begin{figure}
\begin{centering}
\includegraphics[width=1\columnwidth]{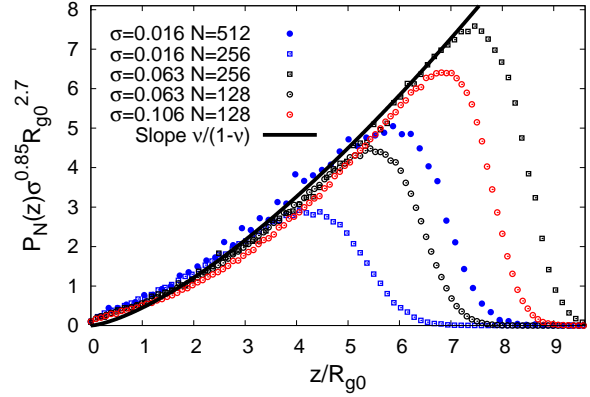}
\par\end{centering}

\caption{\label{figendsswollen}Scaling of the distribution of free ends $P_{N}(z)$
as following the present approach.}
\end{figure}

In the following, we use the normalization $\int_{0}^{H}P_{N}(z)dz=1$
to work with a universal $P_{N}(z)H=f(z/H)$. At small $z$, resp.\ small
$y=z/H$ in eqs.\ (\ref{gegeneral}) and (\ref{gegauss}), one can
approximate the chain end distribution as $P_{N}(z)H\propto y^{\nu/(1-\nu)}$,
according to the present approach, or as $P_{N}(z)H\propto y$, according
to 'standard' SSA. Since $H\propto N\sigma^{(1-\nu)/2\nu}$ (resp.\ $H\propto N\sigma^{1/3}$)
and $N^{\nu}\propto R_{g0}$ (the radius of gyration of a single free
chain in good solvent), we transform both expressions for the chain
end distributions into a universal form as function of $R_{g0}$.
Within the present approach we predict $P_{N}(z)$ to show universality
at small $z$ if presented as 
\begin{equation}
P_{N}(z)\,\sigma^{1/(2\nu)}R_{g0}^{(1+\nu)/\nu}=f(z/R_{g0})\propto\left(\frac{z}{R_{g0}}\right)^{\nu/(1-\nu)}.\label{scalingswollen}
\end{equation}
For $\nu\approx0.588$ this becomes 
\begin{equation}
P_{N}(z)\,\sigma^{0.85}R_{g0}^{2.7}\approx f(z/R_{g0})\propto\left(\frac{z}{R_{g0}}\right)^{1.43}.\label{scalingswollen2}
\end{equation}
In contrast the 'standard' SSA expects universality for 
\begin{equation}
P_{N}(z)\,\sigma^{2/3}R_{g0}^{7/3}=f(z/R_{g0})\propto\frac{z}{R_{g0}}.\label{scalinggauss}
\end{equation}

\begin{figure}
\begin{centering}
\includegraphics[width=1\columnwidth]{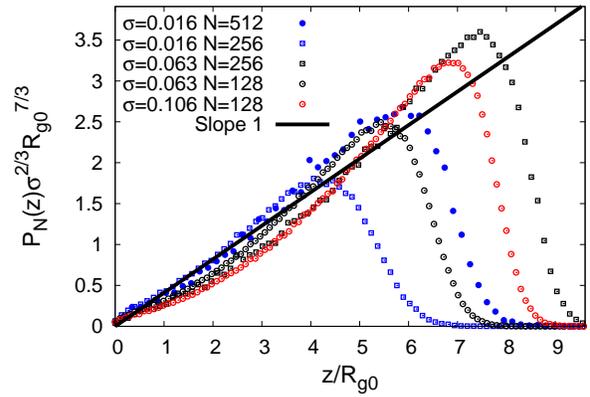}
\par\end{centering}

\caption{\label{figendsgauss}Scaling of the distribution of free ends $P_{N}(z)$
as following the 'standard' approach.}
\end{figure}

In Figures \ref{figendsswollen} and \ref{figendsgauss} we present
our simulation data in the corresponding scaling forms. The radius
of gyration $R_{g0}$ is measured from the BFM simulations of a single
free chain at the corresponding chain lengths. Indeed, at small $z$
in Figure \ref{figendsswollen} the curves roughly fall onto each
other and deviate as the brush height for the corresponding sample
is approached. In contrast, in Figure \ref{figendsgauss} such a universal
behavior cannot be observed at any range of the $z$ position and
the scaling form due to the 'standard' SSA clearly fails. Hence, the
new approach not only accounts for the qualitative difference in form
of a changed slope for $P_{N}(z)$, but it also provides a better
scaling of the simulation data.

\section{Discussion}

\label{Discussion}

In the present work, we combined the model for the elasticity of self-avoiding
chains and the scaling expression for the osmotic pressure in semi-dilute
polymer solutions using a generalized 'Strong Stretching Approximation'
\cite{AmoskovPryam} for the description of polymer brushes. The effect
of self-avoidance of the polymer chains in brushes has not been considered
in literature so far. Our results are fully consistent with scaling
models for the brush height. In the case of good solvent, we predict
deviations from the well-known parabolic density profile towards a
more step-like form and a rather convex shape of the free end distribution
$P_{N}(z)$ at small height values $z$, which is in contrast to the
concave shape obtained previously \cite{MWC,ZhuPriBo}.

Our predictions are supported by simulation data of polymer brushes
at very low grafting densities where finite extensibility can be savely
ignored. Furthermore, our simulation results and the model predictions
are in qualitative agreement with previously observed discrepancies
between simulation data at low grafting densities $\sigma$ \cite{MuratGrest,ChakTor,DickHong,LaiBinder1,LaiBinder2,ChenFwu,HeMerlitz,KaraBit}
and the classical SSA approximation. Since the valid regime of our
approximation - polymer brushes at low to moderate grafting densities
and chain stretchings - is often realized in experimental situations,
we expect that our contribution is significant for further research.
In particular, the predicted changes in the density profile and the
free end distribution are fundamental for understanding the compression
behaviour of brushes, the interpenetration of two opposing brushes,
or the relaxation of chains inside a brush.

\section{Acknowledgement}

We especially thank R.\ Dockhorn and M.\ Werner for providing the
BFM data and kindly wish to thank R.\ Dockhorn, S.\ Egorov, T.\ Kreer,
J.-U.\ Sommer, H. Merlitz, M.\ Werner and E.\ B.\ Zhulina for helpful
discussions. The authors thank the DFG for financial support via the
SPP1369 and the ZIH Dresden for a generous grant of computing time.


\begin{thebibliography}{10}
\bibitem{MWC} S. T. Milner, T. A. Witten, and M. E. Cates, ``Theory
of the grafted polymer brush'', \emph{Macromolecules} \textbf{21},
2610 (1988).

\bibitem{ZhuPriBo} E. B. Zhulina, V. A. Pryamitsyn, and O. V. Borisov,
``Structure and conformational transitions in grafted polymer chain
layers. A new theory'', \emph{Polymer Science USSR} \textbf{31},
205 (1989).

\bibitem{Flory}P. J. Flory, ``The Configurations of Real Polymer
Chains'', \emph{J. Chem. Phys.} \textbf{17}, 303 (1949).

\bibitem{ShimCates} D. F. K. Shim and M. E. J. Cates, ``Finite extensibility
and density saturation effects in the polymer brush'', \emph{Journal
de Physique France} \textbf{50}, 3535 (1989).

\bibitem{LaiHalperin} P.-Y. Lai and A. Halperin, ``Polymer brush
at high coverage'', \emph{Macromolecules}, \textbf{24}, 4981 (1991).

\bibitem{AmoskovPryam} V. M. Amoskov and V. A. Pryamitsyn, ``Theory
of monolayers of non-gaussian polymer chains grafted onto a surface'',
\emph{J. Chem. Soc.Faraday Trans.}, \textbf{90}, 889 (1994).

\bibitem{BieshAmoskov} P. M. Biesheuvel, W. M. de Vos, and V. M.
Amoskov, ``Semianalytical continuum model for nondilute neutral and
charged brushes including finite stretching'', \emph{Macromolecules},
\textbf{41}, 6254 (2008).

\bibitem{Dirk 1}D. Romeis, H. Merlitz, J.-U. Sommer, ''A New Numerical
Approach to Dense Polymer Brushes and Surface Instabilities'', \emph{J.
Chem. Phys.} \textbf{136}, 044903 (2012).

\bibitem{Dirk 2}D. Romeis, J.-U. Sommer, ``Conformational Switching
of Modified Guest Chains in Polymer Brushes'', \emph{J. Chem. Phys.}
\textbf{139}, 044910 (2013).

\bibitem{MuratGrest}M. Murat, G. S. Grest, ``Structure of a Grafted
Polymer Brush: A Molecular Dynamics Simulation'', \emph{Macromolecules}
\textbf{22}, 4054 (1989).

\bibitem{ChakTor}A. Chakrabarti, R. Toral, ``Density Profile of
Terminally Anchored Polymer Chains: A Monte Carlo Study'', \emph{Macromolecules}
\textbf{23}, 2016 (1990).

\bibitem{DickHong}R. Dickman, D. C. Hong, ``New Simulation Method
for Grafted Polymeric Brushes'', \emph{Journal of Chemical Physics}
\textbf{95}, 4650 (1991).

\bibitem{LaiBinder1}P.-Y. Lai, K. Binder, ``Structure and Dynamics
of Grafted Polymer Layers: A Monte Carlo Simulation'', \emph{Journal
of Chemical Physics} \textbf{95}, 9288 (1991).

\bibitem{LaiBinder2}P.-Y. Lai, K. Binder, ``Structure and Dynamics
of Polymer Brushes near the $\Theta$ Point: A Monte Carlo Simulation'',
\emph{Journal of Chemical Physics} \textbf{97}, 586 (1992).

\bibitem{ChenFwu}C.-M. Chen, Y.-A. Fwu, ``Monte Carlo Simulations
of Polymer Brushes'', \emph{Physical Review E} \textbf{63}, 011506
(2000).

\bibitem{HeMerlitz}G.-L. He, H. Merlitz, J.-U. Sommer, C.-X. Wu,
``Static and Dynamic Properties of Polymer Brushes at Moderate and
High Grafting Densities: A Molecular Dynamics Study'', \emph{Macromolecules}
\textbf{40}, 6721 (2007).

\bibitem{KaraBit}E. Karaiskos, I. A. Bitsanis, S. H. Anastasiadis,
``Monte Carlo Studies of Tethered Chains'', \emph{Journal of Polymer
Science: Part B: Polymer Physics} \textbf{47}, 2449 (2009).

\bibitem{AlexanderBrush} S. Alexander, ``Adsorption of chain molecules
with a polar head - a scaling description'', \emph{J. Physique (Paris)}
\textbf{38}, 983 (1977).

\bibitem{deGennesBrush} P. G. de Gennes, ``Conformations of polymers
attached to an interface'', Macromolecules \textbf{13}, 1069 (1980).

\bibitem{deGennes_Book}P. G. de Gennes, ``Scaling Concepts in Polymer
Physics'', Cornell University Press, Ithaka (1979).

\bibitem{Pincus}P. Pincus, ``Excluded Volume Effects and Stretched
Polymer Chains'', \emph{Macromolecules} \textbf{9}, 386 (1976).

\bibitem{LeGuillou}J. C. Le Guillou, J. Zinn-Justin, ``Critical
Exponents for the n-Vector Model in Three Dimensions from Field Theory'',
\emph{Phys. Rev. Lett. }\textbf{39}, 95-98 (1977).

\bibitem{Zhulina_PE}E. B. Zhulina, O. V. Borisov, T. M. Birshtein,
``Structure of grafted polyelectrolyte layer'',\emph{ J. Phys. II
France }\textbf{2}, 63-74 (1992).

\bibitem{Flory_Book}P. J. Flory, ``Principles of Polymer Chemistry'',
Cornell University Press, Ithaca, N.Y. (1966).

\bibitem{Wittmer_2004}J. P. Wittmer, H. Meyer, J. Baschnagel, A.
Johner, S. Obhukov, L. Mattioni, M. M�ller, A. N. Semenov, ``Long
Range Bond-Bond Correlations in Dense Polymer Solutions'', \emph{Phys.
Rev. Lett. }\textbf{93}, 147801 (2004).

\bibitem{Wittmer_2007}J. P. Wittmer, P. Beckrich, H. Meyer, A. Cavallo,
A. Johner, J. Baschnagel, ``Intramolecular long-range correlations
in polymer melts: Te segmental size distribution and its moments'',
\emph{Phys. Rev. E }\textbf{76}, 011803 (2007).

\bibitem{Lang_2013}M. Lang, ``Monomer Fluctuations and the Distribution
of Residual Bond Orientations in Polymer Networks'', \emph{Macromolecules}
\textbf{46}, 9782 (2013).

\bibitem{Milner_MM89_2}S.~T. Milner, Z.-G. Wang, and T.~A. Witten,
``End-confined polymers: Corrections to the newtonian limit'', \emph{Macromolecules},
\textbf{22}, 489 (1989).

\bibitem{Milner_JCSFT_90}S.~T. Milner, ``Strong-Stretching and
Scheutjens-Fleer Descriptions of Grafted Polymer Brushes'', \emph{J.
Chem. Soc. Faraday Trans.}, \textbf{86}, 1349 (1990).

\bibitem{Rubinstein} M. Rubinstein and R. H. Colby, ``Polymer Physics'',
Oxford University Press, Oxford (2003).

\bibitem{KreerMetzger}T. Kreer, S. Metzger, M. M�ller, K. Binder,
J. Baschnagel, ``Static Properties of End-tethered Polymers in Good
Solution: A Comparison between Different Models'', \emph{Journal
of Chemical Physics} \textbf{120}, 4012 (2004).

\bibitem{ZhuBoPry}E. B. Zhulina, O. V. Borisov, V. A. Pryamitsyn,
T. M. Birshtein, ``Coil-Globule Type Transitions in in Polymers.
1. Collapse of Layers of Grafted Polymer Chains'', \emph{Macromolecules}
\textbf{24}, 140 (1991).

\bibitem{Lang_Brush}M. Lang, H. Hoffmann, R. Dockhorn, M. Werner,
J.-U. Sommer, ``Fluctuation driven height reduction of crosslinked
polymer brushes: A Monte Carlo study'', \emph{J. Chem. Phys. }\textbf{139},
164903 (2013).

\bibitem{Carmesin}I. Carmesin, K. Kremer, ``The Bond Fluctuation
Method: A New Effective Algorithm for the Dynamics of Polymers in
All Spatial Dimensions'', \emph{Macromolecules} \textbf{21}, 2819
(1988).

\bibitem{DeutschBinder}H. P. Deutsch, K. Binder, ``Interdiffusion
and self-diffusion in polymer mixtures: A monte Carlo Study'',
\emph{J. Chem. Phys. }\textbf{94}, 2294-2304 (1991).

\bibitem{Lang_Brush2}M. Lang, M. Werner, R. Dockhorn, and T. Kreer,
\emph{in preparation}.

\bibitem{Nedelcu}S. Nedelcu, M. Werner, M. Lang, J.-U. Sommer, ``GPU
Implementations of the Bond Fluctuation Model'', \emph{ J. Comp.
Phys.} \textbf{231}, 2811 (2012).\end{thebibliography}
\end{document}